\documentclass[useAMS,usenatbib,usegraphicx]{mn2e}
\usepackage{graphicx}
\usepackage{amssymb}

\def\kpc{\,{\rm kpc}}

\def\Dmin{{\,D_{\rm min}}}

\newcommand{\mc}{\multicolumn}
\def\magsec{\,{\rm mag\,arcsec^{-2}}}

\begin{document}

\title[Ongoing growth of BCGs via major dry mergers]
{Ongoing growth of the brightest cluster galaxies via major dry mergers in the last $\sim6$ Gyr}

\author[F. S. Liu et al.]
{
F. S. Liu$^{1,2}$\thanks{E-mail:fsliu@synu.edu.cn},
F. J. Lei$^{2,3}$,
X. M. Meng$^{3}$,
D. F. Jiang$^{2}$
\\
$^{1}$Center for Theoretical Physics and Astrophysics, Shenyang Normal University, Shenyang 110034, China\\
$^{2}$College of Physical Science and Technology, Shenyang Normal University, Shenyang 110034, China\\
$^{3}$National Astronomical Observatories, Chinese Academy of Sciences, A20 Datun Road, Beijing 100012, China\\
}

\date{Accepted 2014 November 28; Received 2014 November 18; in original form 2014 August 1}

\pagerange{\pageref{firstpage}--\pageref{lastpage}} \pubyear{2014}

\maketitle

\label{firstpage}

\begin{abstract}

Brightest Cluster Galaxies (BCGs) might have been assembled relatively late ($z<1$) via mergers. 
By exploiting the high-resolution HST/ACS imaging, we find four BCGs (COSMOS-P 125516, 102810, 036694 and 089357) in major dry merging
in 29 X-ray clusters at $0.3 \le z \le 0.6$ in the Cosmological Evolutionary Survey (COSMOS). 
These BCGs show prominent but quiescent double nuclei with a magnitude difference of $\delta m<1.5$ and 
a projected separation of $r_p<$ 10 kpc. Clear signatures of interaction such as 
extended plumes and/or significant asymmetries are also observed in their residual images. 
We infer a major merger rate of $0.55\pm0.27$ merger per Gyr at $z\sim0.43$ assuming the merger time-scale 
estimate of \citet[][]{KW08}. This inferred rate is significantly higher than the rate in the local Universe ($0.12\pm0.03$ at $z\sim0.07$) 
presented in \citet[][]{lmd+09}. We estimate that present-day BCGs increase their luminosity (mass) by $\sim35\pm15$ per cent $(f_{mass}/0.5)$ 
via major dry mergers since $z=0.6$, where $f_{mass}$ is the mean mass fraction of companion galaxies accreted onto the central ones. 
Although the statistical uncertainty due to our small sample size is relatively large, our finding is consistent with 
both recent observational and theoretical results. Furthermore, in conjunction with our previous findings in \citet[][]{lmd+09}, 
the discovery of these intermediate-redshift merging BCGs is clear evidence of ongoing assembly of BCGs 
via major dry mergers over the last $\sim$6 Gyr. 

\end{abstract}

\begin{keywords}
galaxies: elliptical and lenticular, cD - galaxies: clusters: general - galaxies: photometry
\end{keywords}

\section{Introduction} \label{intro}
Brightest Cluster Galaxies (BCGs) are among the most luminous and 
most massive galaxies in the Universe. They are usually located close to 
the centres of dense clusters of galaxies as witnessed in the X-ray
or gravitational lensing observations \citep[e.g.,][]{jf84,Smith+05}. 
Except in strong cooling flows, they are dominated by old stars 
and they lack prominent ongoing star formation (von der Linden et al. 2007; Liu et al. 2012a,b).
%\citep[][\liu{revise it}]{vbk+07,Liu+12,lmm12}.
%(von der Linden et al. 2007; Liu et al. 2012a,b). 
It was noted very early on that some BCGs show excess of light (`envelopes') relative to the de 
Vaucouleurs ($r^{1/4}$) profile at large radii \citep[][]{Matthews+64,Schombert88,Graham+96} and 
were thus termed as cD galaxies. Recent studies have identified cD galaxies 
using improved Petrosian \citep[][]{Petrosian76} 
parameter profiles \citep[][]{Brough+05,Patel+06,liu08}. 
A large fraction of massive nearby BCGs can be classified as cD galaxies \citep[][]{liu08}.

In order to understand the formation and evolution of BCGs/cDs, 
several formation mechanisms including galactic cannibalism \citep[][]{White76,OH77}, 
tidal stripping of stars from interacting galaxies in clusters \citep[][]{GO72,Richstone76,Merritt85} and 
star formation by cooling flows onto BCGs \citep[][]{Fabian94} have been proposed in early studies. 
Recent numerical simulations and semi-analytic models of galaxy formation suggested that 
BCGs form in a two-phase process: an initial collapse with rapid cooling and star formation 
at high redshift is followed by later ($z<1$) growth through multiple dissipationless (dry) 
mergers of pre-existing progenitors \citep[e.g.,][]{db07,RS09,Naab+09,Laporte+12,Laporte+13}. 
These studies, however, differ in their predictions of stellar 
mass growth rates and/or the role of mergers. For instance, \citet[][]{db07} predict that the stellar mass of BCGs increase by a 
factor of $\sim4$ between redshift $z=1$ and $z=0$ via accretion of smaller galaxies (minor mergers). 
More recent simulations in a $\rm {\Lambda}CDM$ universe in \citet[][]{Laporte+13} show 
that BCGs can form through dissipationless mergers of quiescent massive $z = 2$ galaxies and 
predict a lower stellar mass growth for BCGs by both 
major and minor mergers, with mass growth of a factor of 2.1 over the redshift interval $z\sim1.0-0.3$ and 
a further factor of $\sim$1.4 between $z=0.3$ and the present day. 

Observationally, a vast number of studies over the last few years provide evidence that 
BCGs have experienced dry mergers \citep[e.g,][]{bhs+07,Lauer+07,vbk+07,liu08} 
and have built up a large part of their stellar mass via mergers 
at $z<1$ \citep[e.g,][]{lmd+09,Liu+13,TA12,EP12,Lidman+12}. 
The estimated mass growth factors in these studies \citep[][]{Lidman+13,Burke+13,Ascaso+14} 
range from $\sim1.8$ to $\sim2.5$, suggesting major mergers as the primary formation mechanism. 

However, contrary to the above observations and the most recent simulations \citep[e.g,][]{Laporte+13}, 
there are other studies that advocate for little change in stellar masses and sizes of BCGs since $z=1$ \citep[e.g,][]{Collins+09,Stott+10,scb+11}. 
These studies compared high-redshift BCGs in X-ray clusters to their counterparts 
in present-day clusters of the same X-ray luminosity. They claimed that the X-ray selected systems at $z\sim1$ already have more than $90\%$ of their final stellar mass 
and $\sim70 \%$ of their final sizes. 
A possible interpretation of this discrepancy could be a selection bias \citep[][]{Lidman+13}, 
or an early relaxation of high-redshift BCGs in X-ray clusters, which would therefore imply that
they change little in their subsequent evolution.  

Direct observational evidence for dry merging has indeed been found in low-redshift ($z<0.3$) 
central galaxies in groups and clusters by a number of researchers 
\citep[e.g,][]{McIntosh+08,lmd+09,Rasmussen+10,Brough+11,EP12}. 
Some evidence has also been presented at higher redshift ($z>0.3$). 
For example, \citet[][]{Mulchaey+06} and \citet[][]{Jeltema+07} reported 
some examples of dry mergers involving central galaxies in intermediate-redshift groups. 
\citet[][]{tmg+08} reported a observational analysis of supergroup SG 1120-1202 at 
$z\sim0.37$, which is expected to merge and form a cluster as massive as to Coma cluster. 
They argued that group environment is critical for the process of major dry merging. 
\citet[][]{rfv07} reported a very massive cluster (CL0958+4702) at $z=0.39$, 
in which a major dry merger is ongoing to build up the more massive 
BCG. They argued that major mergers might be fairly common in deep 
images of galaxy clusters at intermediate redshifts, and BCGs may be assembled 
through late major mergers. \citet[][]{Lidman+13} identified three merging pairs 
in their high-redshift cluster sample, only the pair in RDCX 1252 ($z=1.238$) shows evidence for a merger. 

In the local Universe, it is known that the incidence rate of major mergers involving central galaxies in clusters 
\citep[][]{lmd+09} is higher than that in groups \citep[][]{McIntosh+08}. In this work, we perform a search for 
ongoing major mergers in a homogeneous sample of intermediate-redshift 
clusters. We find four BCGs (COSMOS-P 125516, 102810, 036694 and 089357) in major dry merging 
in 29 X-ray-selected clusters at $0.3 \le z \le 0.6$ 
in the Cosmological Evolutionary Survey (COSMOS) by \citet[][]{fgh+07}.
The high-resolution HST/ACS F814W band imaging data reveals the prominent morphological signatures 
of dry mergers (close and comparable multiple nuclei, broad surrounding plumes and/or 
significant asymmetries in the isophotes) in these systems. We studied their stellar population, 
investigated possible connection of mergers with the formation of BCG envelopes, and discussed 
the stellar assembly of BCGs in the last $\sim6$ Gyr.
Throughout the paper we adopt a cosmology with a matter density parameter $\Omega_{\rm m}=0.3$, 
a cosmological constant $\Omega_{\rm \Lambda}=0.7$ and 
a Hubble constant of ${\rm H}_{\rm 0}=70\,{\rm km \, s^{-1} Mpc^{-1}}$

\section{Identifications of BCGs and Major mergers}\label{iden}

We searched for BCGs involved in major mergers in a catalog 
of X-ray clusters published by \citet[][hereafter F07]{fgh+07}, which 
includes 72 X-ray clusters at $z_p<1.3$ identified from the first 36 XMM-Newton pointings 
on the $\sim1.7$ $\rm deg^2$ COSMOS field. 
We restricted our search in 29 X-ray clusters in the redshift interval of $z=0.3-0.6$. 
It is difficult to identify dry merger features, such as broad stellar `fans' and 
diffuse tails \citep[][]{Rix+89,Combes+95}, at higher redshift with HST/ACS F814w band images in COSMOS 
because of the significant effects of cosmic surface brightness dimming and bandpass shifting.
Despite the small size of this sample, it is a more statistically homogeneous sample compared to 
other samples collected from many small HST snapshot programs, since COSMOS 
is the largest contiguous field ever imaged with HST. 

\begin{figure*}
\centering
\includegraphics[angle=0,width=1.0\textwidth]{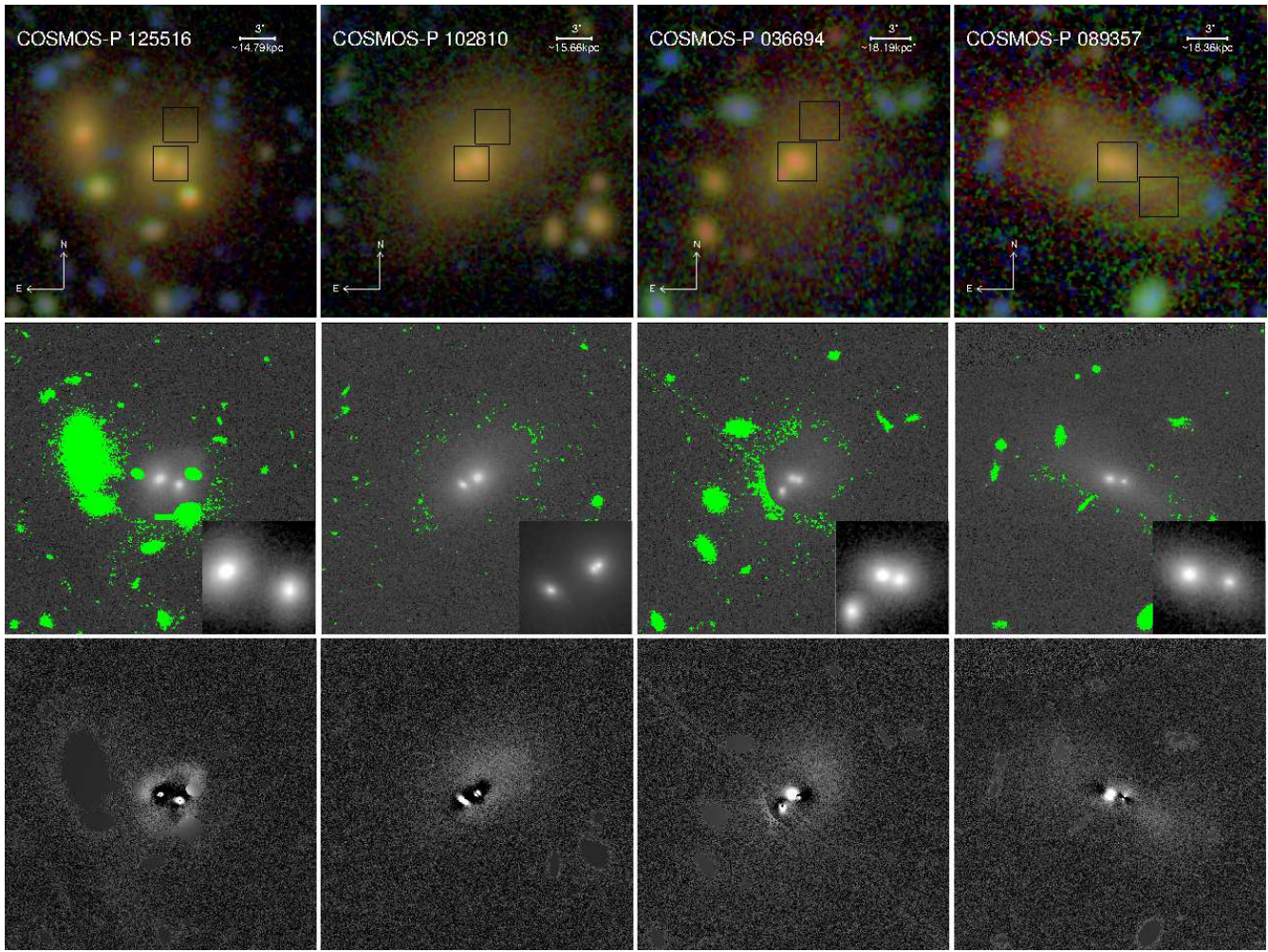}
\caption{ The composite pseudo-color images made with the Subaru
$B_J$,$V_J$,$i^+$-band images (top), HST/ACS F814W band images
(middle) and corresponding residual images (bottom) for four identified BCGs in major merging, respectively.
The small black boxes ($3\arcsec \times 3\arcsec $) in each pseudo-color image mark the central and
outer plume regions analyzed in \S3. The masked regions (green) in the
HST/ACS F814W band images during our modeling procedure are also
shown. The central region ($3\arcsec \times 3\arcsec $ box) are
shown in logarithmic $min/max$ scale on the right-bottom corner of the HST/ACS
F814W band images, which can reveal the spectacular multiple
nuclei more clearly. The centers are determined by averaging the outer isophotes
between the measured $R_e$ and $2R_e$ (see \S4). In each image, north is up and east is to the left.
} \label{fig:samp}
\end{figure*}

\begin{table*}
\scriptsize
%\tiny
%\footnotesize
\caption{Basic parameters for four identified BCGs in major dry merging.}
\setlength{\tabcolsep}{0.012in}
\begin{center}
\begin{tabular}{c|c|c|c|c|c|c|c|c|c|c|c|c|c|c|c|c}
\hline

\mc{1}{c}{Cluster ID} &
\mc{1}{c}{$L_{0.1-2.4 keV}$} &
\mc{1}{c}{BCG ID} &
\mc{1}{c}{ R.A.(J2000) } &
\mc{1}{c}{Dec.(J2000) } &
\mc{1}{c}{$z_p$ } &
\mc{1}{c}{$z_s$ } &
\mc{1}{c}{$M_{F814W}$ } &
\mc{1}{c}{$R_e$ } &
\mc{1}{c}{$M_{\ast}$} &
\mc{1}{c}{$r_p$ } &
\mc{1}{c}{$t_{merge}$ } &
\mc{1}{r}{$m_{fit}$} &
\mc{1}{c}{$m_{res}$ } &
\mc{1}{c}{$f_{res}/f_{ini}$ } &
\mc{1}{c}{${\cal A}(1.5\arcsec)$ } &
\mc{1}{r}{${\cal A}(2R_e)$ }  \\

\mc{1}{c}{(1)} & \mc{1}{c}{(2)} & \mc{1}{c}{(3)} & \mc{1}{c}{(4)} & \mc{1}{c}{(5)} & \mc{1}{c}{(6)} & \mc{1}{c}{(7)} & \mc{1}{c}{(8)} & \mc{1}{c}{(9)} & \mc{1}{c}{(10)} & \mc{1}{c}{(11)} & \mc{1}{c}{(12)} & \mc{1}{c}{(13)} & \mc{1}{c}{(14)} & \mc{1}{c}{(15)} & \mc{1}{c}{(16)} & \mc{1}{c}{(17)} \\

\mc{1}{c}{} & \mc{1}{c}{($10^{42}$ erg $s^{-1}$)} & \mc{1}{c}{} & \mc{1}{c}{} & \mc{1}{c}{} & \mc{1}{c}{} & \mc{1}{c}{} & \mc{1}{c}{(mag)} & \mc{1}{c}{(kpc)} & \mc{1}{c}{($M_{\odot}$)} & \mc{1}{c}{(kpc)} & \mc{1}{c}{(Gyr)} &\mc{1}{c}{(mag)} & \mc{1}{c}{(mag)} & \mc{1}{c}{} & \mc{1}{c}{} & \mc{1}{c}{}\\

\hline
52  & $3.1\pm0.4$ & COSMOS-P 125516 & 150.37774 & 2.41224 & 0.3489 & 0.3488 & -23.86 & 6.55  & $1.22\times10^{11}$ &  & 0.32  & 18.24 & 19.49  & 16.1 & 2.71 & 1.35   \\
    &             &             & 150.37727 & 2.41210 &        &        &        &       &                    & 8.75   &    & 18.61 &        &      &      &       \\
64  & $3.0\pm0.3$ & COSMOS-P 102810 & 150.23333 & 2.47626 & 0.3816 &        & -24.05 & 7.04  & $1.84\times10^{11}$ &  & 0.23   & 18.55 & 19.34  & 18.6 & 2.97 & 1.56  \\
    &             &             & 150.23336 & 2.47623 &        &        &        &       &                    & 0.63   &    & 20.38 &        &      &      &       \\
    &             &             & 150.23368 & 2.47608 &        &        &        &       &                    & 7.15   &    & 18.51 &        &      &      &       \\
120 & $8.0\pm0.9$ & COSMOS-P 036694 & 149.75633 & 2.79472 & 0.4917 & 0.4936 & -24.24 & 7.21  & $2.49\times10^{11}$ &  & 0.23   & 17.87 & 19.99  & 16.4 & 2.87 & 1.42  \\
    &             &             & 149.75621 & 2.79469 &        &        &        &       &                    & 2.74   &    & 19.01 &        &      &      &       \\
    &             &             & 149.75656 & 2.79446 &        &        &        &       &                    & 7.55   &    & 20.34 &        &      &      &       \\
80  & $7.3\pm1.1$ & COSMOS-P 089357 & 150.11266 & 2.55611 & 0.5022 &        & -24.09   & 12.91 & $1.73\times10^{11}$ & & 0.22   & 18.30 & 19.94  & 19.9 & 1.56 & 0.25  \\
    &             &             & 150.11237 & 2.55605 &        &        &        &       &                    & 6.51   &    & 19.66 &        &      &      &       \\

\hline
\hline
\end{tabular}
\end{center}
\label{tab:bcg}
{Note: Col:(1) Cluster ID in the F07 cluster catalog.
Col:(2) Rest-frame luminosity in the 0.1-2.4 keV band of the cluster in the F07 cluster catalog.
Col:(3) BCG ID from the NASA/IPAC Extragalactic Database (NED).
Col:(4) R.A.(J2000.0) of member (nucles) in merging. 
Col:(5) Dec.(J2000.0) member (nucleus) in merging. 
Col:(6) Photometric redshift of the BCG. 
Col:(7) Spectroscopic redshift of the BCG. 
Col:(8) Absolute magnitude in HST/ACS F814W I$-$band. No k-correction and extinction correction are applied.  
Col:(9) Effective radius of the whole merging system.
Col:(10) Stellar mass estimated by the SED fitting. 
Col:(11) The projected separation of members (nuclei) in merging. 
Col:(12) The merger time-scale estimated by the formula of Kitzbichler \& White (2008)
Col:(13) The best fitted magnitude for each member (or nucleus) by GALFIT.
Col:(14) The magnitude of residual images after subtracting the best models. The masked region is not included in the calculaton. 
Col:(15) The flux ration of redidual and initial image within $3R_e$. The masked region is not included in the calculaton.    
Col:(16) The asymmetric factor (${\cal A}$) within $3\arcsec$ aperture ($1.5\arcsec$ radius).    
Col:(17) The asymmetric factor (${\cal A}$) within $2R_e$. 
The parameters in Col.(8)-(10) are from the measurements on the regions brighter than 26 $\magsec$. 
}
\end{table*}

The F07 X-ray cluster catalog provides the center and the photometric redshift of each cluster candidate, 
but it does not provide the BCG candidate in each cluster. The cluster center is defined by the peak of X-ray emission. 
The majority of photometric redshifts are estimated using the early photo-$z$ catalog of 
galaxies in COSMOS \citep[][]{Mobasher+07} and the rest of the photometric redshifts are from other sources 
(see F07 for details). The accuracy of photo-$z$ in the catalog of \citet[][]{Mobasher+07}
is ${\sigma}_{\Delta z}/(1+z) \sim 0.027$. Subsequently, \citet[][]{Ilbert+09} improved 
the photo-$z$ accuracy by using 30-bands photometric data and provided absolute magnitudes 
in the Subaru $V_J$-band for galaxies in the $\sim$2 $\rm deg^2$ COSMOS field. 
The accuracy of photo-$z$ is improved to ${\sigma}^{'}_{\Delta z}/(1+z) \sim 0.012$. 
Since the used photometric redshifts in F07 have larger uncertainties, 
in order to select reliable BCG candidates, we firstly selected galaxies 
within the radius of 1 Mpc from the cluster center and a photo-$z$ gap of 
$z\pm3{{\sigma}_{\Delta z}}$ as temporary member candidates in a cluster. 
We re-calculated the cluster redshift to be the median value of photometric redshifts 
of the recognized `members' using the more accurate photo-$z$ values in \citet[][]{Ilbert+09}. 
We then applied this newly-derived cluster redshift to re-select member candidates 
within the radius of 1 Mpc from a given cluster center and a photo-$z$ gap of $z\pm1.44{{\sigma}^{'}_{\Delta z}}$, 
and re-calculated the absolute magnitudes of these member galaxies. In this step, 
we used the spectroscopic redshifts instead of photo-$z$ if they were available 
from the zCOSMOS redshift survey \citep[][]{Lilly+09}. 
We also supplemented the photometric redshifts for X-ray sources from \citet[][]{Salvato+09}. 
The brightest one in the member galaxies in a cluster is finally selected as our BCG candidate.

We then extracted the imaging data of 29 sample BCGs from COSMOS Archive 
website\footnote{http://irsa.ipac.caltech.edu/data/COSMOS/} and performed visual inspection. 
Four BCGs (COSMOS-P 125516, 102810, 036694 and 089357) in the corresponding clusters 
with ID of 52, 64, 120, 80 in the F07 catalog caught our high attention 
due to significant merging features, such as close multiple nuclei and extended surrounding plumes, 
shown in the high-resolution HST/ACS F814W band images. 
Their composite pseudo-color images made with the Subaru $B_J$,$V_J$,$i^+$-band images (top panels) 
and HST/ACS F814W band images (middle panels) are shown in the Figure~\ref{fig:samp}, respectively. 
Some characteristics of these candidates are listed as follows, ordered in increasing 
redshift (hereafter the same ordering will be adopted). Note that there are the spectroscopic redshifts available 
for COSMOS-P 125516 and 036694 only. 
We used the photometric redshifts available in \citet[][]{Ilbert+09} 
for the other two candidates. 

COSMOS-P 125516 is a typical pair of early-type galaxies with a
projected separation of 1.77$\arcsec$ ($\sim8.75\kpc$). The merging
features (e.g., bridges and plumes) are obvious.

COSMOS-P 102810 looks like a merging `pair' of early-type
galaxies, but the high-resolution HST/ACS imaging reveals spectacular
double nuclei in one of members, which have a projected separation of
only 0.12$\arcsec$ ($\sim0.63\kpc$). Broad surrounding plumes
to the northwest of its image can be seen clearly, extending to $>9$$\arcsec$ ($\sim47\kpc$, also see \S4).

COSMOS-P 036694 is a triple system. The two closest nuclei have a
projected separation of only 0.45$\arcsec$ ($\sim2.74\kpc$).
Broad surrounding plumes, extending to $>10$$\arcsec$ ($\sim60\kpc$), is also obvious on the northwest of the BCG. 

COSMOS-P 089357 has close double nuclei with a projected
separation of 1.06$\arcsec$ ($\sim6.51\kpc$). It has very broad
and symmetric stellar plumes extending to $>10$$\arcsec$ ($\sim61\kpc$) along the semi-major axis.

We further quantified our visual inspection with a method used in previous studies \citep[][]{wlh09,lmd+09} to 
efficiently identify merging signatures in the absence of spectroscopic information for merging members. We applied the GALFIT package \citep[][]{phi+02} to
construct a smooth symmetric S{\'e}rsic (S{\'e}rsic 1968) model 
for every target member in the projected merging list. The modeling procedure gives a fitted
magnitude for each target member and a residual image for each candidate. 

All these candidates have at least a pair of members (nuclei)
with the magnitude difference $\delta m_{fit}<1.5$, which can be classified as major mergers
\citep[][]{bnm+06}. All of them have significant residuals
and merging signatures, which are seen more clearly in their
residual images (see bottom panels of Figure~\ref{fig:samp}). The
magnitudes of the four residual images are 19.49, 19.34, 19.99 and 19.94,
which correspond to 16.1\%, 18.6\%, 16.4\% and 19.9\% of their initial
images, respectively. The asymmetric factor (${\cal A}$) for each
residual can be calculated following the method of
\citet[][]{wlh09}. This factor measures the brightness difference between
pixels and those of symmetric pixels with respect to centers of target members.
The derived ${\cal A}$ values for COSMOS-P 125516, 102810, 036694, and 089357 
are 2.71, 2.97, 2.87 and 1.56 for their inner
regions within 3$\arcsec$ aperture (1.5$\arcsec$ radius), respectively. The ${\cal A}$
values become 1.35, 1.56, 1.42 and 0.25 for the regions within twice measured 
effective radius ($2R_e$). It shows that these candidates have significant asymmetry in their inner and/or 
outer isophotes \citep[e.g.,${\cal A}>0.45$,][]{wlh09}. Note that the outer isophotes of COSMOS-P 089357 
are relatively symmetric, indicating this merging system has been well-relaxed in its outskirts and 
a perfect structure resembling cD galaxy has formed (see \S4). Some derived
parameters for four candidates are listed in Table~\ref{tab:bcg}.

\section{Stellar population analysis}\label{SED}

\begin{figure}
\centering
\includegraphics[angle=0,width=0.5\textwidth]{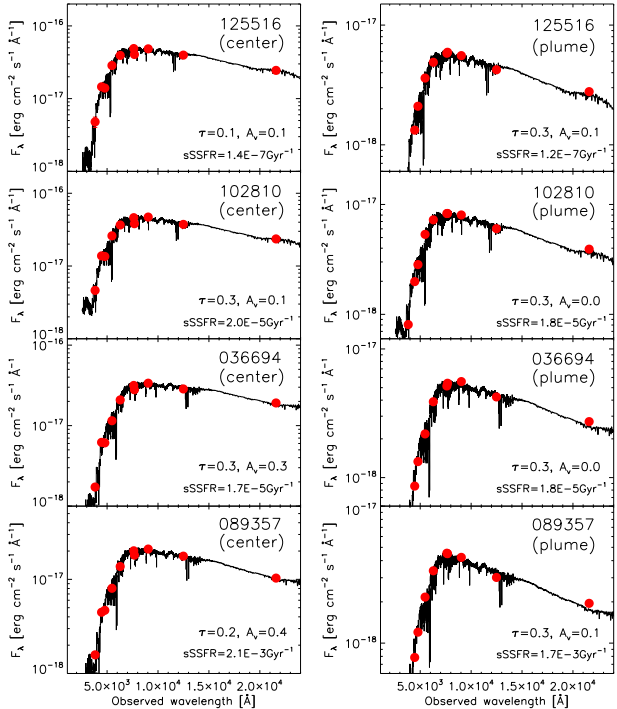}
\caption{The best fittings on the observed SEDs in the regions of
centers (left panels) and outer plumes (right panels) of COSMOS-P
125516, 102810, 036694 and 089357 (from top to bottom), respectively.
Red dots are the observed data. The best-fitted spectra are from
a series of evolutionary CSPs models constructed following
\citet[][]{bkp+}. The $e$-folding time ($\tau$) shown on bottom right
corner of each panel is the closest value that the target region has in
the exponentially declining SFHs. The derived specific star formation rate (sSFR)
and $A_V$ for each region are also presented.
The sizes of observed data dots are larger than their error bars.
\label{fig:seds}}
\end{figure}

We investigated stellar population of four major mergers by fitting their spectral
energy distributions (SEDs) in optical and near-infrared
broad bands measured from multi-wavelength imaging data available in COSMOS
survey. We used the CFHT($u*$, $i*$ and $K_s$-band), Subaru
($B_J$, $V_J$, $g^+$, $r^+$, $i^+$ and $z^+$-band) and UKIRT
($J$-band) observations because they are also deep and with high
spatial resolution (0.15$''$/pixel). Since it has been shown that 
the evolutionary composite stellar populations (CSPs) can 
more exactly describe the evolutionary history of COSMOS galaxies \citep[][]{bkp+}, 
we followed \citet[][]{bkp+} to construct a series of evolutionary 
CSP templates from the libraries of simple stellar populations (SSPs) of \citet[][hereafter BC03]{bc03} 
with exponentially declining star formation histories (SFHs), 
with $e$-folding times $\tau$=0.1, 0.2, 0.3, 0.5, 1 and 2 Gyr. 
We used an updated version of the code $Hyperz$ \citep[][]{bmp00}, kindly made available to us by M.
Bolzonella, which performs SED fitting at a fixed redshift. We
performed the best fittings on the observed SEDs in the regions of
centers and outer plumes within 3$"$ boxes (marked on the top
panels of Figure~\ref{fig:samp}) with the fixed solar metallicity
and Chabrier initial mass function \citep[][]{bkp+}. The dust extinction was modeled
using the Calzetti extinction law \citep[][]{cab+00}. 
The results (see Figure~\ref{fig:seds}) show the SEDs of both centers and outer plumes of four systems 
can be well described by an exponentially declining SFH with a very short $e$-folding time (0.1 - 0.3 Gyr) 
and with extremely low specific star formation rates (sSFRs $\rm <10^{-2} Gyr^{-1}$).
The results from the SED fitting indicate that the star formation history of these four BCGS is consistent with that of 
typical quiescent elliptical galaxies.

\begin{figure}
\begin{center}
\begin{minipage}[t]{0.95\linewidth}
\centerline{\includegraphics[angle=0,width=0.99\textwidth]{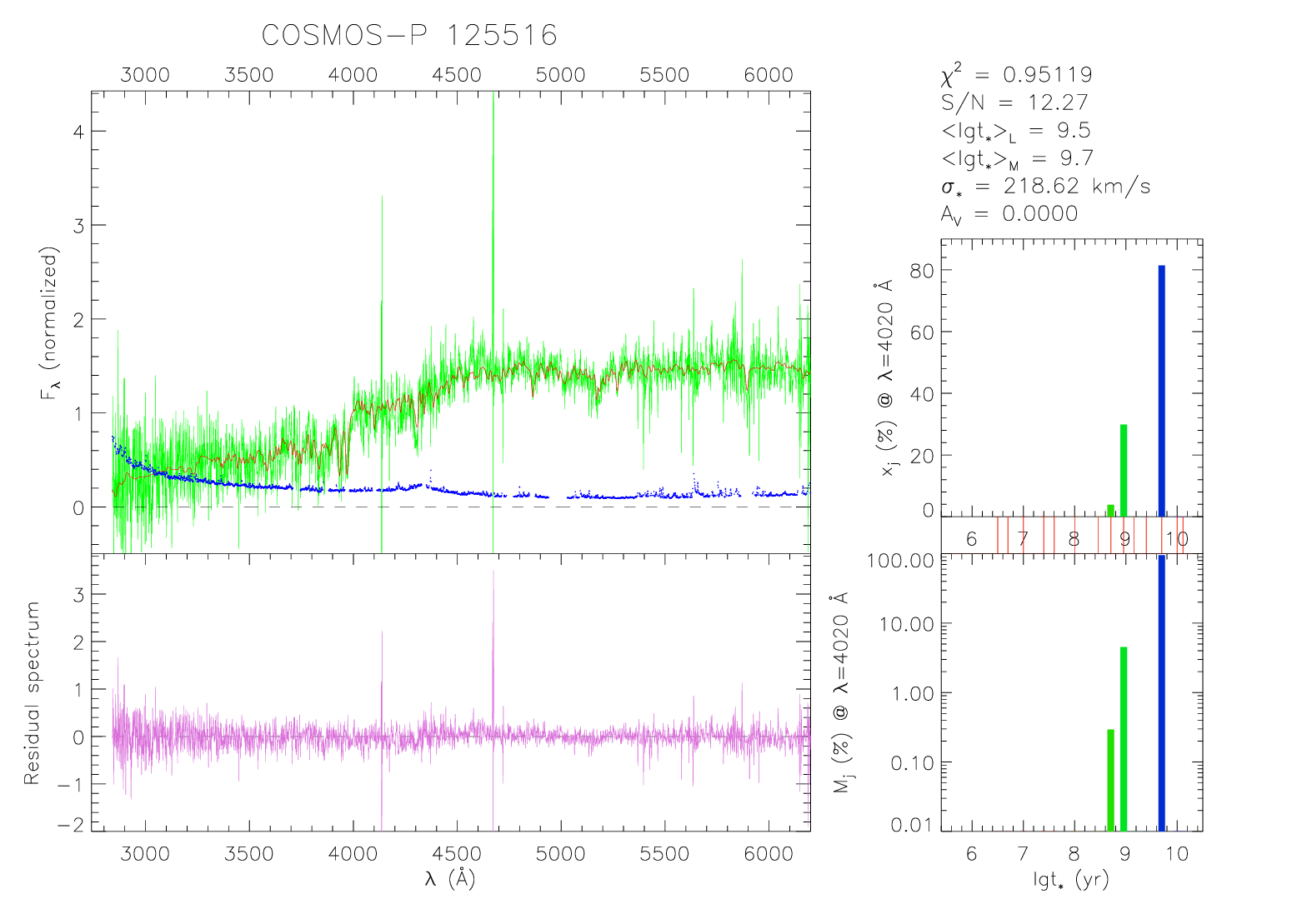}}
\end{minipage}%
\\
\begin{minipage}[t]{0.95\linewidth}
\centerline{\includegraphics[angle=0,width=0.99\textwidth]{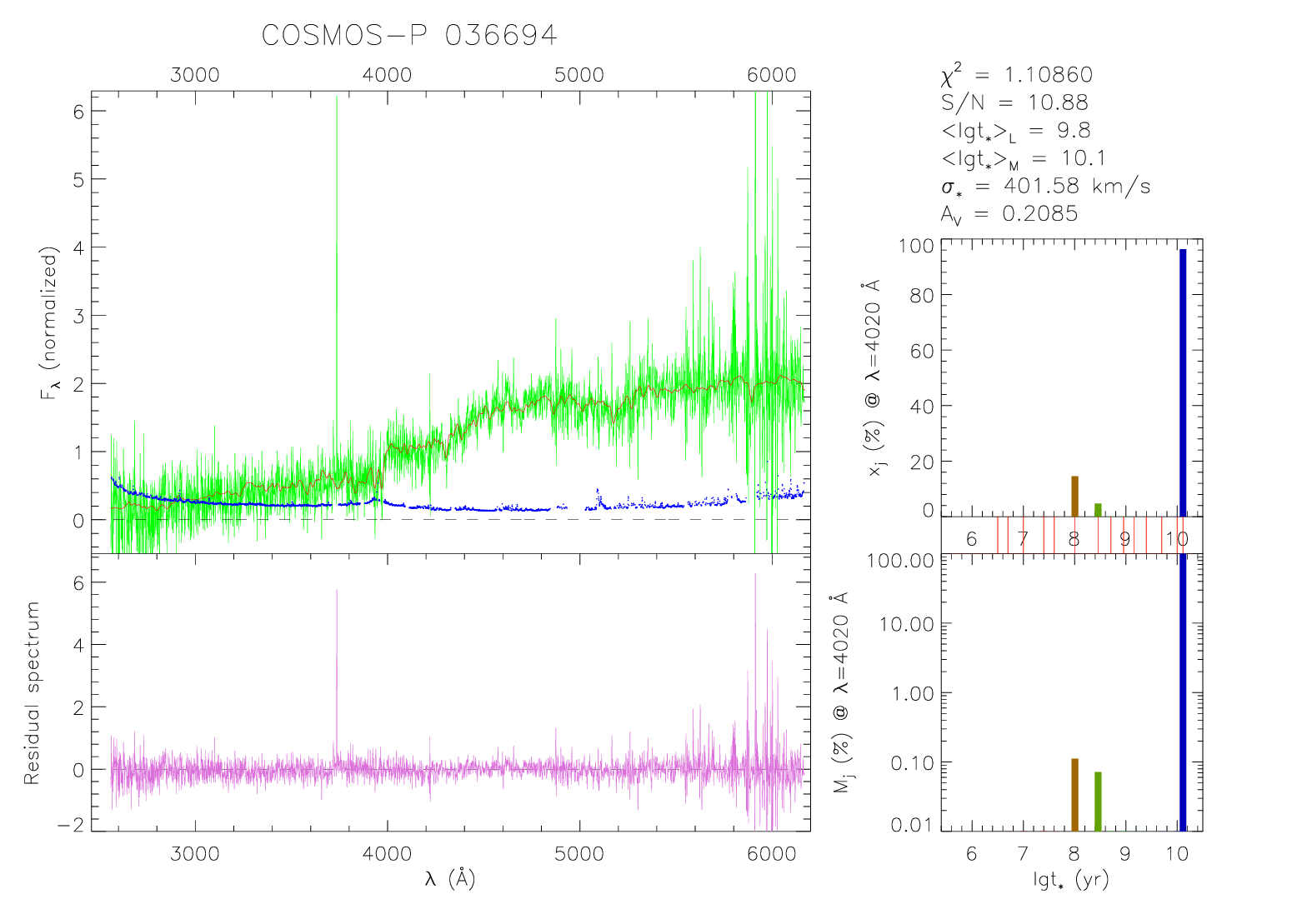}}
\end{minipage}%
\end{center}
\caption{
The spectral synthesis for COSMOS-P 125516 (top) and 036694 (bottom), respectively.
The observed spectra (green) and  model spectra (red) are shown in left panels.
The residual spectrum is shown on the bottom left panel for each object.
The flux intensities in the left panels are normalized at 4020{\rm \AA}\, by $4.5\times10^{-16}\,
{\rm (ergs\ s^{-1}\ cm^{-2})}$. The light and mass weighted stellar population fractions
are shown in the right panels. Some derived quantities by the fitting are also presented.
The S/N ratio is measured in the S/N window between 4730 and 4780 {\rm \AA}.
} \label{fig:sepc}
\end{figure}

The Sloan Digital Sky Survey (SDSS) has spectroscopic data available for COSMOS-P 125516 and 036694, 
which have the signal-to-noise ratio (S/N) of 12.27 and 10.88, respectively. 
Since their SFHs can be well modeled by an exponentially declining SFR with very short e-folding
timescale ($\tau \le 0.3$) that is close to an instantaneous burst
($\tau = 0 $). We follow \citet[][]{lmm12} to adopt a serial
of spectral templates of SSP with the fixed solar metallicity but
different ages to estimate the light and mass percentage ratios of
their stellar populations at different ages. We show that the light and
mass ratio among young ($t<0.5$ Gyr), intermediate age ($0.5<t<2.5$ Gyr)
and old ($t>2.5$ Gyr) stellar population is 3.37 : 25.95 : 70.68 and 0.29 : 4.51 : 95.20 for COSMOS-P 125516, 
and 16.73 : 0 : 83.27 and 0.18 : 0 : 99.82 for COSMOS-P 036694, respectively. It shows recent star
formation only contributes a very small (even negligible) percentage to their stellar mass. This result is roughly consistent with the finding in our early study of low-redshift BCGs \citep[][]{lmm12}. 

\section{Envelopes \& merging process \label{sec:envelope}}

\begin{figure*}
\centering
\includegraphics[angle=0,width=1.0\textwidth]{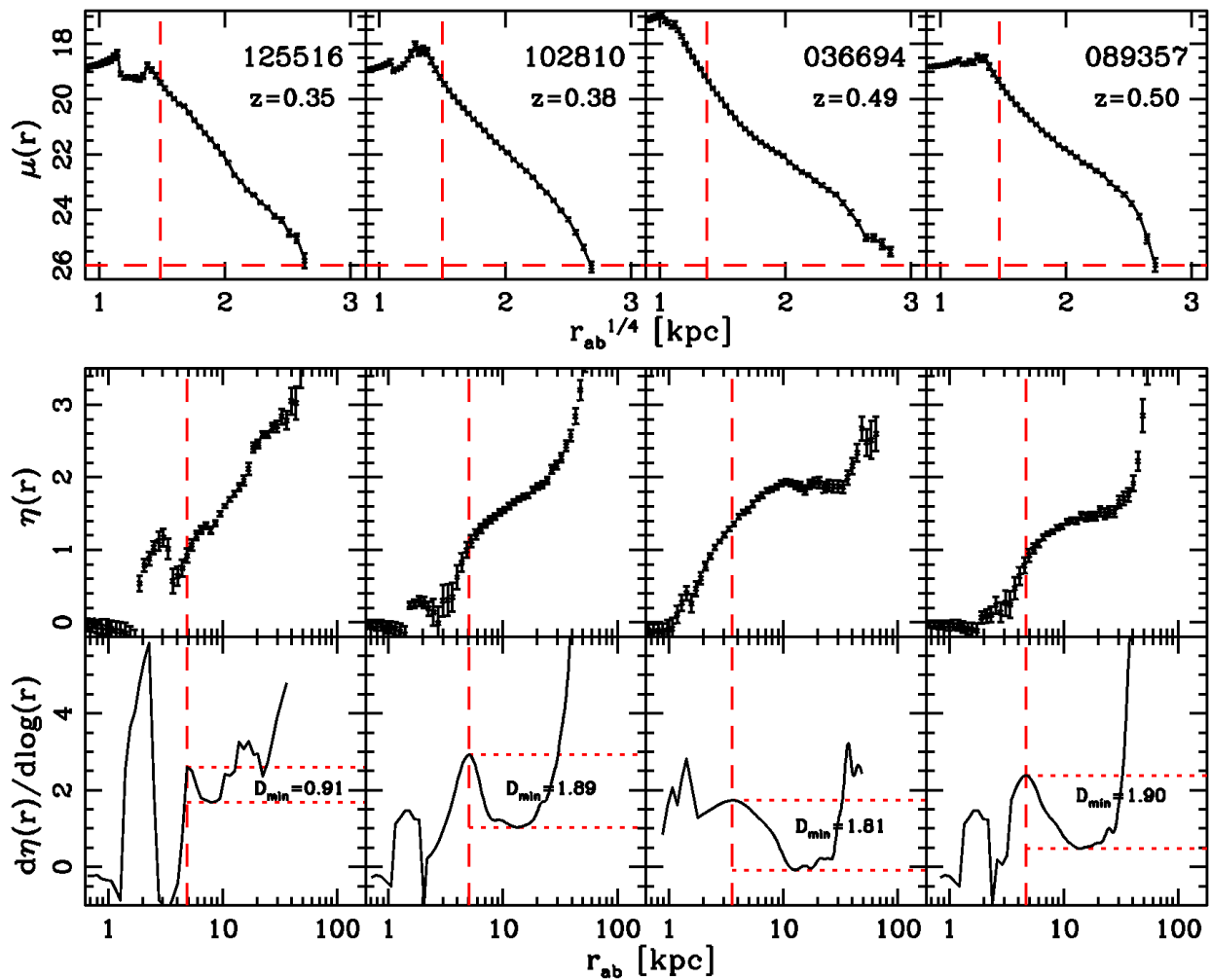}
\caption{
The HST/ACS F814W band surface brightness profiles, Petrosian $\eta(r)$ profiles and $d\eta(r)/d\log(r)$ profiles
for four merging BCGs, respectively. The surface brightness profiles have been corrected for cosmic dimming.
The equivalent radius of an ellipse, $r_{ab} \equiv \sqrt{ab}$, is used, where
$a$ and $b$ are the lengths of semi-major and semi-minor axes of the best-fitted ellipse.
The red horizontal lines in the first rows show the surface brightness of 26 $\magsec$.
The red vertical line in each panel marks the radius, within which the profiles fluctuate significantly due to
the existence of multiple nuclei. The depths of valleys in $d\eta(r)/d\log(r)$ profiles are indicated with red horizontal dotted lines and
corresponding $\Dmin$ values are presented.
\label{merger_spec.eps}}
\end{figure*}

The surface brightness profiles of a large fraction of local BCGs 
show strong deviation from a perfect de Vaucouleurs (de Vaucouleurs 1948) or
S$\acute{\rm e}$rsic (S$\acute{\rm e}$rsic 1968) profile 
since these BCGs are embedded in extensive luminous stellar halos. 
The envelopes (halos) in BCGs can be recognized more clearly by identifying the signatures of
plateau in the Petrosian $\eta(r)$ profiles and valley in the
$\gamma(r)=d\eta(r)/d\log(r)$ profiles \citep[][]{liu08}. 
We have presented evidence that the formation of extended stellar envelopes in local BCGs 
are closely connected with dry merger process by analyzing the gradient of 
their Petrosian profiles \citep[see][for details]{liu08}.
Here we investigate if such envelopes are observed in these merging systems at intermediate redshift. 
We performed the surface photometry on these four systems following the method used in \citet[][]{liu08}.  
Figure~\ref{merger_spec.eps} shows their surface brightness profiles 
to $26\magsec$ in HST/ACS F814W band, corresponding Petrosian $\eta(r)$ profiles 
and $\gamma(r)$ profiles from top to bottom, respectively. It can be seen that these parameter profiles 
fluctuate significantly in inner regions (within the radius marked with red vertical dashed lines 
in Figure~\ref{merger_spec.eps}) 
due to the existence of multiple nuclei. However, their light profiles at large radii are relatively smooth. 
There are obvious plateaus in the $\eta(r)$ profiles and valleys in the $\gamma(r)$ profiles of all four mergers. 
The $\Dmin$ values derived in the $d\eta(r)/d\log(r)$ profiles of four systems (COSMOS-P 125516, 102810, 036694, and 089357) 
are 0.91, 1.89, 1.81, and 1.90 (see bottom panels in Figure~\ref{merger_spec.eps}), respectively. 
These values are larger than the median $\Dmin$ value ($\sim 0.8$) for local BCG sample in \citet[][]{liu08}.
It shows that they have extended envelopes in their outskirts and are likely forming cD galaxies, moreover, 
the broad stellar fans can also be seen clearly on their images (see bottom panels of Figure 1). 
The signatures of cD galaxies in both intermediate-redshift and local merging BCGs \citep[][]{lmd+09} 
and the fans corresponding to cD galaxy envelopes suggest that the extended stellar halos of BCGs 
are likely due to major mergers and they appear shortly after merger takes place.

\section{Major merger rate and Stellar assembly of BCGs} \label{discussion}

We have identified 4 BCGs in major dry merging in 29 X-ray clusters 
at $0.3 \le z \le 0.6$ (with a median value of $z\sim0.43$). We estimate the merger 
time-scale for each merger by the equation (1) in \citet[][]{lmd+09}, which is derived from 
the Millennium Simulation by \citet[][]{KW08}. We find that the merger time-scale 
of our mergers ranges from 0.22 to 0.32 Gyr with a mean value of 0.25 Gyr. 
(see Column 11 in Table 1). Thus, we obtain a major dry merger rate of $0.55\pm0.27$ merger 
per Gyr at $z\sim0.43$, where the error is assumed to be a Poisson error. In our previous work \citep[][]{lmd+09}, 
we used a broader criterion to select 18 ($\sim3.5$ percent of the total) major dry merging 
pairs (or triples) involving the BCG in 515 C4 clusters at $0.03 \le z \le 0.12$ 
(with a median value of $z\sim0.07$). The time-scale of those local mergers 
ranges from 0.04 to 0.55 Gyr with a mean value of 0.3 Gyr \citep[see][for details]{lmd+09}. 
A major merger rate of $0.12\pm0.03$ at $z\sim0.07$ can thus be determined as well. 
Parameterizing the evolution of major merger rate in the form of $(1+z)^m$, we find that 
$m=5.2\pm2.4$ at $z \le 0.6$. Here, we stress that the number of BCGs in our intermediate-redshift sample 
is small and the resulting statistical uncertainty is relatively large. 
The two predominant members (nuclei) in our intermediate-redshift mergers have an 
average magnitude difference of $\sim0.79$, which is comparable to that ($\sim0.7$) 
of local merger pairs/triples in \citet[][]{lmd+09} and corresponds to a $\sim1:2$ 
luminosity ratio. 
We assume a fraction of $f_{mass}$ of the companion galaxy is accreted to the primary galaxy. 
Therefore, in each merger the central galaxy increases its luminosity 
(mass) by 25 percent ($f_{mass}/0.5$) on average. 
If the major merger rate evolves in our estimated form above since $z=0.6$ and 
the merger time-scale follows the formula of \citet[][]{KW08}, it follows 
that a present-day BCG has on average increased its mass (luminosity) 
by $\sim35\pm15$ per cent ($f_{mass}/0.5$) from $z=0.6$ at a mean rate of $6\pm2.6$ per cent 
($f_{mass}/0.5$) per Gyr. Thus, a large fraction of the stellar mass of a present-day 
BCG is assembled via major dry mergers in the last $\sim6$ Gyr.

\section{Summary \& Discussion} \label{discussion}

In this work, we have identified four BCGs in major dry merging in 29 X-ray clusters 
at $0.3 \le z \le 0.6$, which are selected in a homogeneous cluster sample in COSMOS 
published by \citet[][]{fgh+07}. These major mergers have two predominant members (nuclei) 
with the magnitude difference of $\delta m<1.5$ and projected separation of $r_p<$ 10 kpc, 
and showing signatures of interaction in the form of broad stellar plumes and/or 
significant asymmetries in residual images. 
They are composed of an old, passively evolving stellar population,
and with negligible amount of young stars.
Photometric analysis show that broad stellar plumes 
in their outskirts have roughly formed extended stellar envelopes (cD halos), 
which provides evidence for the connection of mergers with the formation of stellar halos. 

We have obtained a major dry merger rate of $0.55\pm0.27$ merger per Gyr at $z\sim0.43$ 
if the time-scale of our mergers follows the calibration of \citet[][]{KW08}.
This major merger rate at moderate redshift is higher than 
that rate of $0.12\pm0.03$ at $z\sim0.07$ presented in \citet[][]{lmd+09}. 
We estimate that the major merger rate may increase in the form of $(1+z)^{5.2\pm2.4}$ to $z=0.6$. 
We conclude that a present-day BCG has on average increased its luminosity (mass)
by $\sim35\pm15$ per cent (by a factor of $\sim1.5$) via major dry mergers 
from $z=0.6$ at a mean merger rate of $6\pm2.6$ per cent 
per Gyr, under the assumption that half of the mass of the companion is accreted to the primary galaxy ($f_{mass}=0.5$). 
While the statistical uncertainty, which stems from the small size of our intermediate-redshift sample, 
is relatively large, the estimated amount of mass growth is roughly consistent with the most recent predictions 
of \citet[][]{Laporte+13} from numerical simulations. 
\citet[][]{Lidman+13} found that 3 of the 14 BCGs at $z\sim1.1$ are likely to experience a major 
merger within 600 Myr and derived a mass growth rate of 7 per cent under the same assumption. 
Our rate is comparable to their mass growth rate at high redshift. 
\citet[][]{EP12} showed that BCGs at $z\sim0.3$ are adding as much as 10 per cent of their stellar mass 
via both major and minor mergers over 0.5 Gyr. It is not unreasonable that our rate is 
considerably lower than theirs since we include only major mergers in our analysis. 
Our study supports the notion that major dry mergers play the dominant role in the late mass assembly of BCGs. 
In conjunction with our previous findings in \citet[][]{lmd+09}, the discovery of these intermediate-redshift merging BCGs is clear evidence of ongoing assembly of BCGs
via major dry mergers over the last $\sim$6 Gyr.

\section*{Acknowledgments}
We thank Shude Mao, David C. Koo, Hassen M. Yesuf for useful comments, and 
especially thank M. Bolzonella for his code and helpful suggestions. 
We also acknowledge the anonymous referee for a constructive report that improved the paper.
This project was supported by the NSF grants of China (11103013, 11203033) and 
the Program for Liaoning Excellent Talents in University (LNET). 

%\bibliographystyle{mn2e}
%\bibliography{merger}

\label{lastpage}
\end{document}